\documentclass[prl,twocolumn,groupedaddress,amssymb]{revtex4}
\usepackage{graphicx}
\usepackage{bm}

\begin{document}
\title{A direct measurement of the Bose-Einstein Condensation universality class in NiCl$_2$-4SC(NH$_2$)$_2$ at ultra-low temperatures}
\author{L. Yin$^1$}
\author{J.S. Xia$^1$}
\author{V.S. Zapf\,$^2$}
\author{N.S. Sullivan$^1$}
\author{A. Paduan-Filho$^3$}
\affiliation{$^1$Department of Physics, University of Florida,
 and National High
Magnetic Field Laboratory, Gainesville, Florida 32611-8440}.
\affiliation{$^2$National High Magnetic Field Laboratory, Los Alamos
National Laboratory, MS E-536, Los Alamos, NM 87545.}
\affiliation{$^3$Institut de Fisica, Univeridade de Sao Paulo, Brazil.}

\date{\today}

\begin{abstract}
In this work, we demonstrate field-induced Bose-Einstein
condensation (BEC) in the organic compound
NiCl$_2$-4SC(NH$_2$)$_2$ using AC susceptibility measurements down
to 1 mK. The Ni $\textsl{S}$=1 spins exhibit 3D $\mathit{XY}$
antiferromagnetism between a lower critical field $H_{c1} \sim$
2~T and a upper critical field $H_{c2} \sim$ 12~T. The results
show a power-law temperature dependence of the phase transition
line $H_{c1} (T) -H_{c1} (0) = aT^{\alpha}$ with $\alpha =$ 1.47$\pm$0.10
and $H_{c1} (0) = 2.053$ T, consistent with the 3D Bose-Einstein
Condensation universality class. An abnormal change was found
in the phase boundary near $H_{c2}$ at approximately 150~mK.
\end{abstract}
\maketitle

The idea of Bose-Einstein condensation (BEC) occurring in the spin
systems of certain quantum magnets with axial symmetry has been
explored extensively in the past few years~\cite{Giamarchi08}.
This idea was first suggested by Affleck in 1991~\cite{Affleck91}
and first investigated experimentally in the compound TlCuCl$_3$~\cite{Ruegg03}. Although BEC in TlCuCl$_3$ was later called into question in
an oft-overlooked electron spin resonance (ESR) study~\cite{Glazkov04}, the work on TlCuCl$_3$ generated a flurry of
experimental and theoretical activity in the field and many new
candidate BEC systems have since been proposed~\cite{Jaime04,PaduanFilho04,Zapf06,Waki05,Aczel08,Bergman06,Kitada07,Garlea07,Radu05}.

In these compounds, the axial symmetry of the spins allows $XY$
antiferromagnetic order to occur over certain ranges of
temperature and magnetic field. Near the critical magnetic fields
where the long-range order is induced or suppressed, the spin
system can be mapped onto a system of dilute hard-core bosons on a
lattice and the field-induced quantum phase transition at the
boundary of the long-range ordered state can be modeled as a
Bose-Einstein condensation~\cite{Giamarchi99,Batista01,Batista04,Ng05,Wang05}. The caveat is
that in quantum magnets the boson number is proportional to the
longitudinal magnetization, and thus is conserved in equilibrium
rather than strictly on all time scales. Therefore, in quantum
magnets only the $\it{thermodynamic~properties}$ of the system should
follow the predictions of BEC theory and nonequilibrium
effects such as supercurrents will not occur. Nevertheless, these
compounds provide an important test of BEC phase transition in the thermodynamic limit.

A key experimental signature of BEC is a power-law
temperature-dependence of the number of condensed bosons with an
exponent of 3/2. This power-law is the low-temperature limit of
the boson distribution function. In the spin systems this
translates to a power-law temperature dependence of the critical
field line $H_c - H_c(0) \propto T^{\alpha}$ where $\alpha = 3/2$~\cite{Nikuni00,Nohadani04}. {\bf It is important to note} that this power-law is
valid in the limit of very low temperatures. So far very few quantum
magnet BEC candidate have been studied at temperatures well below
the energy scales for boson interactions, which are given by the
antiferromagnetic couplings between spins. Furthermore, if the temperature range at which
the power-law fit is performed is far from zero temperature, then it is very difficult
to accurately identify the intercept, and the resulting power-law
exponent that is derived from the fit is highly dependent on the
value of the intercept used~\cite{Sebastian06b}. One way to circumvent this
problem is the windowing method in which the
intercept and the exponent are determined more-or-less
independently by performing fits over different temperature ranges
and extrapolating the values of the intercept and the exponent to zero temperature~\cite{Sebastian05}. However, this extrapolation technique from
higher temperatures does not always correspond to the actual
low-temperature behavior as was seen in the compound
BaCuSi$_2$O$_6$, where at temperatures above 1 K, the windowing
method yields one exponent, but due to a reduction in
dimensionality of the spin system, a different exponent can be
observed at lower temperatures~\cite{Sebastian06c}.

Here we report a direct observation of the 3/2 power law exponent
of the 3D BEC universality class at ultra low temperatures in the compound DTN (NiCl$_2$-4SC(NH$_2$)$_2$). Our measurements were performed at
temperatures down to 1~mK, which is two orders of magnitude below
the lowest temperature scale for magnetic coupling in this system
$J_a$ = 180~mK, and the lowest temperature ever used to investigate
BEC in a quantum magnet. Thus we do not have to use any extrapolation to determine the power-law exponent.

The organic magnet DTN contains $S$ = 1 Ni$^{2+}$ atoms that form two interpenetrating
tetragonal lattices. At zero field, a uniaxial anisotropy $D \sim
9$~K splits the Ni $S$ = 1 triplet into a $S_z = 0$ ground state and
a $S_z = \pm 1$ excited doublet. The $S_z = 1$ state can be
suppressed with applied magnetic fields along the tetragonal
\textsl{c}-axis via the Zeeman effect, thus producing a magnetic ground
state above $H_{c1} = 2.1$~T. Antiferromagnetic coupling between
the Ni atoms with exchange strength $J_c = 1.8$~K along the \textsl{c}-axis
and $J_a = 0.18$ K perpendicular to the \textsl{c}-axis produces long-range
antiferromagnetic order~\cite{Zvyagin07}. The long-range order occurs in a
dome-shaped region of the
$T-H$ phase diagram between $H_{c1}$,
where the magnetic ground state is induced, and $H_{c2}$ where the
spins align with the applied magnetic field and below $T = 1$~K~\cite{Zapf06}.
The $XY$ symmetry, the magnetic exchange couplings and the uniaxial
anisotropy $D$ have been identified via inelastic neutron
scattering and electron spin resonance (ESR). The power-law
dependence of the critical field line $H_{c1}$ was previously investigated using specific heat,
magnetocaloric effect and magnetization measurements down to a
minimum temperature of 100~mK~\cite{Zapf06,PaduanFilho08} using
the windowing method, and a power-law consistent with $\alpha = 3/2$
was determined. The validity of this power-law was called into question however due to the fact that it is an extrapolation rather a direct measurement in particular in light of the recent new results for BaCuSi$_2$O$_6$ showing a different exponent at low temperatures. \cite{Sebastian06c}

We have now measured the power-law temperature
dependence of the critical field $H_{c1}$ down to 1~mK using AC
susceptibility measurements. The experiments were carried out using
 a PrNi$_5$ nuclear refrigerator and a
15~T magnet at the High $B/T$ facility of National High Magnetic
Field Laboratory.
The sample was immersed in liquid $^3$He in a polycarbonate cell,
and thermal contact to the refrigerator was assured via  sintered  silver  that was an integral part of an assembly of  annealed silver rods extending form the nuclear refrigerator.  The temperature was calibrated by a $^3$He
melting-pressure curve thermometer mounted in the
zero-field region of the magnet, at the top of the nuclear stage.
Both AC and DC magnetic fields were applied in the direction
parallel to the $\textsl{c}$-axis of the single crystal of DTN. An AC
signal with an amplitude of 0.5~$\sim$~1~G and a low frequency
of 275~Hz was generated in a primary coil wound from
superconducting NbTi wire, thereby avoiding heating at ultra-low
temperatures. The AC susceptibility $\chi_{ac}$ was measured by
sweeping the external DC field at rates between 0.1 and$\sim$~0.001~T/min while the
temperature was fixed at various values between 270 and 1.0~mK.

\begin{figure}
\centering
\includegraphics[scale=1.35]{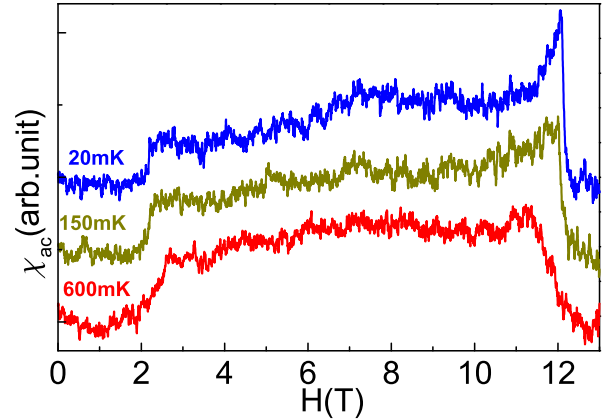}
\caption{(Color online) AC susceptibility, $\chi_{ac}$, as a function
of magnetic field $\mathit{H}$ with a field sweep rate of
0.054~T/min. Traces are shown for $\mathit{T}$=20~mK (the highest
curve), 150~mK and 600~mK.} \label{fig:Figure1}
\end{figure}

Fig.~\ref{fig:Figure1} shows three traces of AC susceptibility
$\chi_{ac}$ for temperatures of 20, 150, and 600~mK with a relatively fast sweep
rate of 0.054~T/min. The typical field-induced long-range AFM (BEC) transitions
appears as steps in these traces with $H_{c1} \sim 2$~T and $H_{c2} \sim 12$~T. As the temperature is lowered, the steps become sharper and a peak develops near $H_{c2}$.

\begin{figure}
\centering
\includegraphics[scale=1.15]{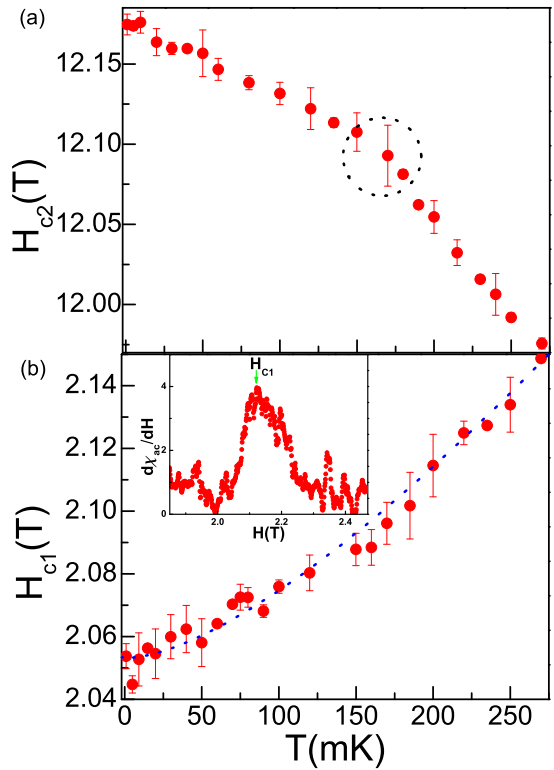}
\caption{(Color online) (a) The temperature dependence of the
upper critical field $H_{c2}$ {\bf for} a field sweep rate of
0.0068~T/min. The dashed circle shows an abnormal change in slope.
(b) The temperature dependence of the lower critical field
$H_{c1}$ in a field sweep rate of 0.0068~T/min. The line is a fit
to the equation $H_{c1}(T) - H_{c1}(0) \sim T^{\alpha}$ with $\alpha
= 1.47 \pm 0.1$ and $H_{c1}(0) = 2.053$~T. Inset to (b): The
temperature dependence of ${\mathrm d}\chi_{ac} / {\mathrm d}H$, with
the critical field $H_{c1}$ indicated by an arrow.}
\label{fig:Figure2}
\end{figure}

The values of $H_{c1}$ and $H_{c2}$ were determined from the peak
in the first derivative of the AC susceptibility ${\mathrm d}\chi_{ac}
/ {\mathrm d}H$, as shown in the inset of Fig.~\ref{fig:Figure2}. In the critical field - temperature phase
diagram shown in Fig.~\ref{fig:Figure2}(a) and Fig.~\ref{fig:Figure2}(b), the temperature dependence of $H_{c1}$ and
$H_{c2}$ have been plotted separately. The data points in Fig.~\ref{fig:Figure2} were collected from the magnetization traces
with a field sweep-rate of 0.0068~T/min. As shown in Fig.~\ref{fig:Figure2}(a), $H_{c2}$ saturates at 12.175~T as
$\mathit{T}$ approaches zero. However, there exists a region
marked by a dashed circle at approximately 150~mK, where a shoulder appears. The
origin of this anomaly is unclear at this point.

\begin{figure}
\centering
\includegraphics[scale=0.46]{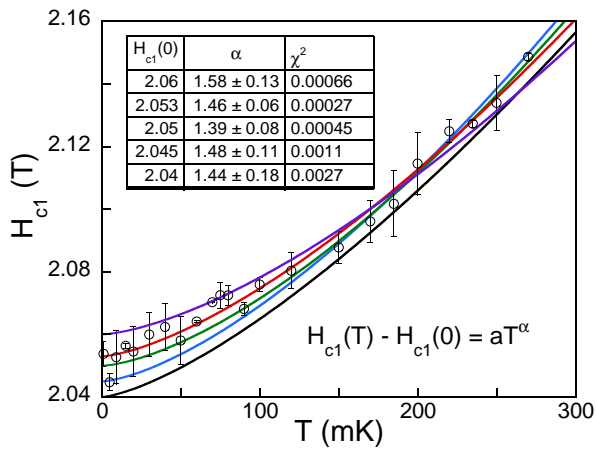}
\caption{(Color online) (a) The critical field $H_{c1}$ as a
function of temperature $T$. The lines are fits of the equation
$H_{c1}(T) - H_{c1}(0) = aT^\alpha$, where $H_{c1}(0)$ is held
fixed and $a$ and $\alpha$ are allowed to vary.}
\label{fig:Figure3}
\end{figure}

\begin{figure}
\centering
\includegraphics[scale=1.10]{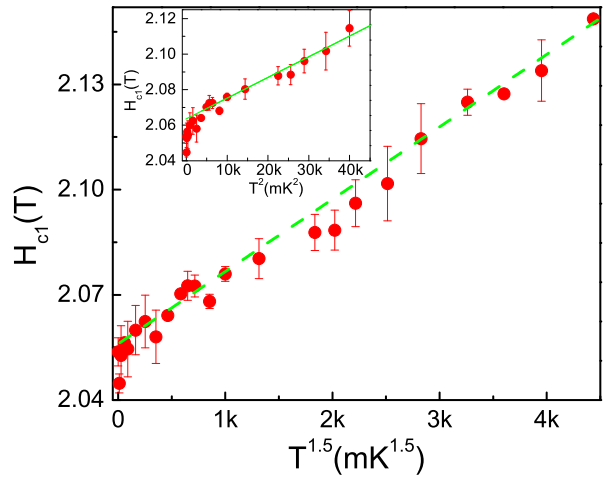}
\caption{(Color online) The lower critical field $H_{c1}$ as a
function of $T^{\alpha}$, where $\alpha \sim 1.5$. The dashed
straight line is the linear fit to this trace. Inset: $H_{c1}$
as a function of $T^{\alpha}$, where $\alpha \sim 2$.}
\label{fig:Figure4}
\end{figure}

For $H_{c1}$ in Fig.~\ref{fig:Figure2}(b), we fit a power-law
temperature dependence $H_{c1}(T) - H_{c1}(0) = aT^{\alpha}$ to
the data between 1 and 260~mK, yielding an intercept of 2.053~T
and an exponent of $1.47 \pm 0.1$ as shown in Fig.
\ref{fig:Figure1}. In order to demonstrate that the exponent is
robust and independent of the intercept, we also performed
fits to the data with the intercept held fixed for various
intercepts between 2.04 and 2.06~T in Fig. \ref{fig:Figure3}.
The table in the figure indicates the exponent, it's error bar,
and the value of $\chi^2$ for each fit. The intercept with the
lowest value of $\chi^2$ is $H_{c1} (0) = 2.053$~T, yielding an exponent
$\alpha$ of 1.46 in that fit. Furthermore, all the fits for all
the intercepts yield values close to 1.5. Finally we show $H_{c1}$
as a function of $T^{1.5}$ in Fig. \ref{fig:Figure4}. The inset
shows the same data as a function of $T^{2}$, showing that we can
rule out that temperature dependence. Thus the best fit yields an
exponent $\alpha = 1.47 \pm 0.1$, which closely matches the expected
exponent $\alpha = 1.5$ for a quantum phase transition in the 3D
BEC universality class. Other nearby universality classes such as
the 3D Ising ($\alpha = 2$) and 2D BEC ($\alpha = 1$) can be
excluded.


Finally, we found that DTN reached thermal equilibrium at 1~mK in a
few minutes which is short in comparison with other solid samples that we have
measured in the same configuration, including heavy fermion and
two-dimensional electronic gas samples with  cooling time constants as
long as hours~\cite{xia00}. DTN as well as the other samples
mentioned here were immersed in the liquid $^3$He, rather than being
glued to a cold finger. There are several possible explanations for
the short thermal equilibrium time at low temperatures. Firstly, the
loose structure of DTN may enhance the cooling efficiency. The
$^3$He atoms can penetrate into the interlayer distance of
8.981~{\AA}. Secondly, there exist possible exchange interactions
between the spins of the $^3$He atoms and electrons in the sample.
Finally, a compatibility between phonon vibration modes in the
$^3$He and the sample may play a role.

In conclusion, we have established that the field-induced quantum
phase transition at $H_{c1}$ in DTN belongs to the 3D BEC
universality class by directly measuring the power-law exponent in
the relation $H_{c1}(T) - H_{c1}(0) \propto T^{\alpha}$ down to 1 mK. This is the first example of a direct measurement of this exponent at temperatures far below the energy scales for antiferromagnetic coupling in a magnetic insulator.

This work was carried out at the High $B/T$ facility of the National High Magnetic Field Laboratory supported by the National Science Foundation Cooperative Agreement No. DMR 0654118 and the State of Florida.

\bibliography{DTN}

\end{document}